\documentstyle[prl,aps,epsf]{revtex}
\draft
\begin{document}
\twocolumn[\hsize\textwidth\columnwidth\hsize\csname @twocolumnfalse\endcsname

\title{Fluctuating Topological Defects in 2D Liquids: Heterogeneous Motion and Noise}
\author{C. Reichhardt and C.J. Olson Reichhardt} 
\address{ 
Center for Nonlinear Studies, Theoretical, and Applied
Physics Divisions, 
Los Alamos National Laboratory, Los Alamos, New Mexico 87545}

\date{\today}
\maketitle
\begin{abstract}
We measure the defect density as a function of time at different temperatures
in simulations of a two dimensional system of interacting particles.
Just above the solid to liquid transition temperature, the
power spectrum of the defect fluctuations shows a $1/f$ signature, which 
crosses over to a white noise signature at higher
temperatures. When $1/f$ noise is present, the 5-7 defects predominately
form string like structures, and the particle trajectories show a 1D
correlated motion that follows the defect strings. At higher temperatures this 
heterogeneous motion is lost. We demonstrate this 
heterogeneity both in systems
interacting with a short ranged screened Coulomb interaction,
as well as in systems with a long range logarithmic interaction
between the particles. 
\end{abstract}
\vspace{-0.1in}
\vspace{-0.3in}

\vskip2pc]
\narrowtext
In liquids and glassy systems there has been considerable
interest in  
dynamical heterogeneities which occur when certain
regions of the sample have a higher mobility than the rest of the sample.   
Particle motion in these systems often occurs by means of
correlated motion of a group of particles along a string like structure.
\cite{Liquieds}.  
Most of the studies of heterogeneities in 2D and 3D 
have focused on systems that are supercooled  near the 
glass transition in simulations \cite{Kob} 
and experiments \cite{Kegel,Cui,Tang}. 
A recent study considered a relatively simple system
of a 2D dense monodisperse 
colloidal assembly which shows an ordered to disordered transition
as a function of density \cite{Cui}. 
In this system the
colloids form a defect free 
triangular lattice at high densities and disorder for lower densities.  
Near the disordering transition at the close packing density,
collective heterogeneous motion of the particles appears,
and the system 
consists of domains of sixfold coordinated particles, $n_c=6$,
surrounded by grain
boundaries composed of strings of
$n_c=7$ and $n_c=5$ dislocations.
The formation of grain boundaries or defect condensation in 
2D monodisperse liquids has also been observed 
in colloidal assemblies \cite{Tang,Marcus} and in dusty plasmas
\cite{Goree,Chiang,Tau}.

Since the dislocations 
have a complex interaction with one another, consisting of a 
long range logarithmic repulsion and a short range attractive 
core interaction, 
one could expect complicated dislocation dynamics for systems with 
even a low density of dislocations.  
In the liquid state, 
creation and annihilation of dislocation pairs
also occurs due to the thermally induced motions of the
underlying particles.  Although there have been numerous studies
of the fluctuations of the particles in 2D liquids,
the fluctuations in the {\it defect density}, defined as 
the number of particles
with $n_c \ne 6$, have not been investigated.
This measure would be easy to access in experiments 
on systems such as colloids and dusty plasmas where  
the individual particles can be directly imaged.
It is not known how the density of $n_c \ne 6$ particles would fluctuate
as the ordering transition is approached. 
If the defects are concentrated in clumps or 
grain boundaries, then it is likely that    
the creation or annihilation of defects will be highly correlated, which 
can give rise to $1/f^{\alpha}$ fluctuation spectra.
Conversely, if the
defects are appearing and disappearing in an uncorrelated manner, 
a white noise spectra would arise.
We also wish to connect the formation of strings of defects with
the appearance of correlated particle motion along 1D strings.
  
In this work we show that for 2D systems which form triangular lattices
at low temperatures, 
for increasing temperature there is a transition from 
a non-defected regime to a defected regime where there is a proliferation
of defects. 
Here we do not attempt to examine the nature of the disordering
transition, such as whether there is an intermediate hexatic
phase \cite{Stranburg}. Instead, we concentrate on the motions 
of the particles and the defect fluctuations in the liquid phase. 
Close to the disordering transition,  in the defected regime, 
the system consists of regions of particles with sixfold order surrounded
by strings or grain boundaries of 5-7 fold defects,
and there are few free dislocations. 
We also find that in this region, 
collective particle motion occurs
in 1D strings along the grain boundaries. 
When temperature is fixed,
the density and structure of the 
defect strings fluctuates with time.
Near the disordering transition,
the
defect density shows large fluctuations with $1/f$ spectra.
For increasing temperatures, the defect density increases 
and the string like structures of the defects and the 1D string like motion 
of the particles are lost. 
At the same time, the low frequency noise power increases and
the spectrum crosses over to a white noise signature,
indicating the
loss of correlations in the creation or annihilation of defects.

We numerically study 2D systems at finite temperature using  
Langevin dynamics.
We consider monodisperse particles in a sample with periodic
boundary conditions and have investigated two types of
interactions.
Most of the results presented here
are for  
particles interacting via a Yukawa or screened Coulomb
interaction potential, 
$V(r_{ij}) = (Q^2/|{\bf r}_{i} - {\bf r}_{j}|)
\exp(-\kappa|{\bf r}_{i} - {\bf r}_{j}|)$. Here ${\bf r}_{i (j)}$ is 
the position of particle $i (j)$, 
$Q$ is the charge of the 
particles, $1/\kappa$ is the screening length, 
and we use 

\begin{figure}
\center{
\epsfxsize=3.4in
\epsfbox{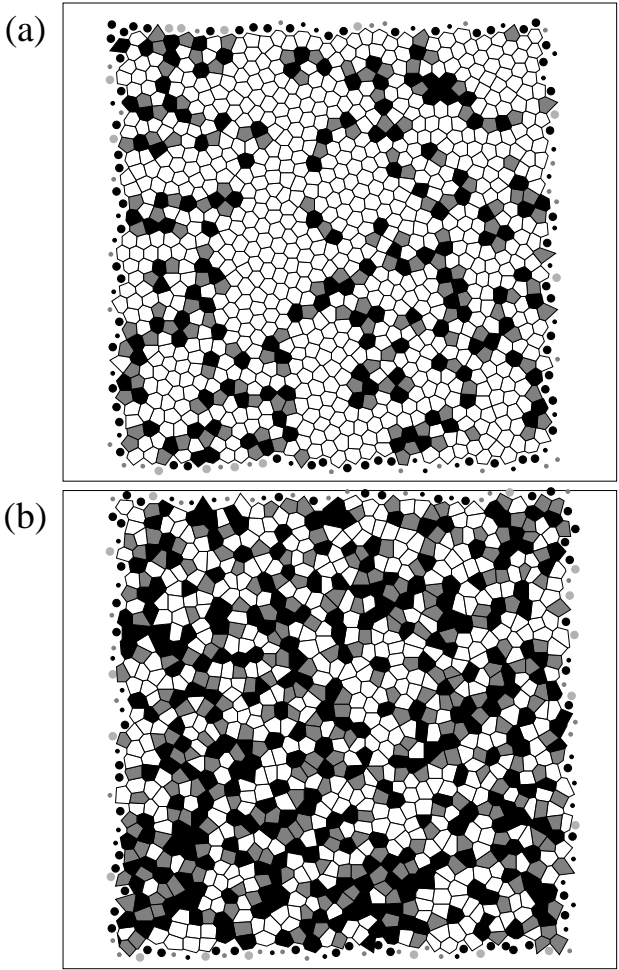}}
\caption{
The Voronoi constructions for a 2D system of particles interacting
with a screened Coulomb potential. A particle is centered at each polygon
and the polygon color is: $n_c=6$, white; $n_c=5$, dark gray,
$n_c=7$, black. (a) 
$T/T_{d} = 1.04$; (b) $T/T_{d} = 7.0$. 
}
\end{figure}

\hspace{-13pt}
$\kappa = 2/a_{0}$, where $a_{0}$ is the lattice constant.
We choose to study this interaction since the quantities we study in this 
work can be accessed in systems such as 2D colloidal assemblies 
and dusty plasmas, in which the
particles interact via a screened Coulomb interaction.
We also consider particles interacting via a
logarithmic potential, $V(r)=\ln(r)$. 
Unlike the Yukawa potential, this interaction is long-range so that 
an interaction cutoff cannot be implemented.  For increased computational
efficiency we use the summation method of Ref. \cite{Jensen}. 
A physical example of logarithmically interacting particles 
with overdamped dynamics is vortices
in thin-film type-II superconductors.

The equation of motion for particle $i$ is:
\begin{equation}
\frac{d{\bf r}_{i}}{dt} = -\sum_{j\neq i}^{N_{c}}\nabla_i V(r_{ij}) 
+ {\bf f}_{T} 
\end{equation}
where 
${\bf f}_{T}$ is a randomly fluctuating force due to thermal
kicks with $<{\bf f}^{T}(t)>=0$ and
$<{\bf f}^{T}(t){\bf f}^{T}(t^{\prime})>=2k_BT\delta(t-t^{\prime})$.    
In all the simulations we 
initially start from an ordered triangular 
configuration. We fix the temperature ${\bf f}_{T}$
for $10^6$ time steps before we begin to take data.
We note that previous simulations have shown that extremely long time
transients can arise near the order to disorder 
transition of up to $2\times 10^6$ 
time steps for system sizes of
$36864$ particles \cite{Canright}. 
In this work we limit ourselves to system sizes of 
$N_{c} = 2000$ 
or less, so that $10^6$ time steps is adequate for equilibration.   
In addition, histograms 
of the time series of the defect density are Gaussian, which
is further evidence that 
our systems are equilibrated. We measure temperature in units of
$T/T_{d}$ where $T_{d}$ is the temperature at which
the first free dislocations
appear.

In Fig.~1(a) we show the Voronoi construction of a 
system of particles with a screened Coulomb interaction 
for $T/T_{d} = 1.04$. The Voronoi construction is similar to the Wigner-Seitz 
cell construction and an individual particle is located in the 
center of each cell. If a particle has six nearest neighbors, $n_c=6$, then 
the 
polygon has six sides. In Fig.~1 particles with $n_c=6$ are white,
$n_c=5$ are dark gray, and $n_c=5$ are black. 
We analyze
a series of such images at fixed $T/T_{d}=1.04$.  We find that
29\% of the particles are defected and that
most (94.5\%) of the defects are part of a cluster or condensate
rather than free.
In some regions, the defects form strings or grain boundaries. 
There are 
also a small number of free disclinations present, indicating
that we are in the liquid phase rather than a hexatic phase, 
according to 2D continuous melting theories.
Clustering of defects
has been observed in experiments on dusty plasmas \cite{Goree} 
which find comparable numbers of free dislocations.   
In Fig.~1(b)  we show the same system 
at a higher temperature, $T/T_{d} = 9$.  Here, the
number of defects is higher
with most of the defects again predominately appearing in 
clusters; however, there is no string like characteristic to the clusters. 
Instead, the defects form clump like objects.

In Fig.~2(a) we show the positions of the particles (black dots) and trajectories
(lines) for the same system in Fig.~1(a). The trajectories are taken for the
particle motion over 10000 time steps. 
Fig.~2(a) shows that  the motion is 
heterogeneous with certain regions in which the particles do not move. These
areas are also the defect free regions. In the areas where there is
motion, the particles move by  $a$, and there is some tendency for the motion
to occur along string like paths. We also often find places 
where the motion occurs in a  circular path with an immobile particle
in the center.
A similar kind of motion in 1D like strings and rotations was seen
in experiments on dense colloidal liquids \cite{Cui}. 
If we take trajectories for longer times, 
eventually all the particles take part in some motion. 
In Fig.~2(b) we show the positions and the trajectories for the 
high temperature phase in system 
Fig.~1(b) for $2000$ time steps. Here the motion occurs everywhere
with no evidence for correlated stationary regions. We find that
even for very short times, 

\begin{figure}
\center{
\epsfxsize=3.5in
\epsfbox{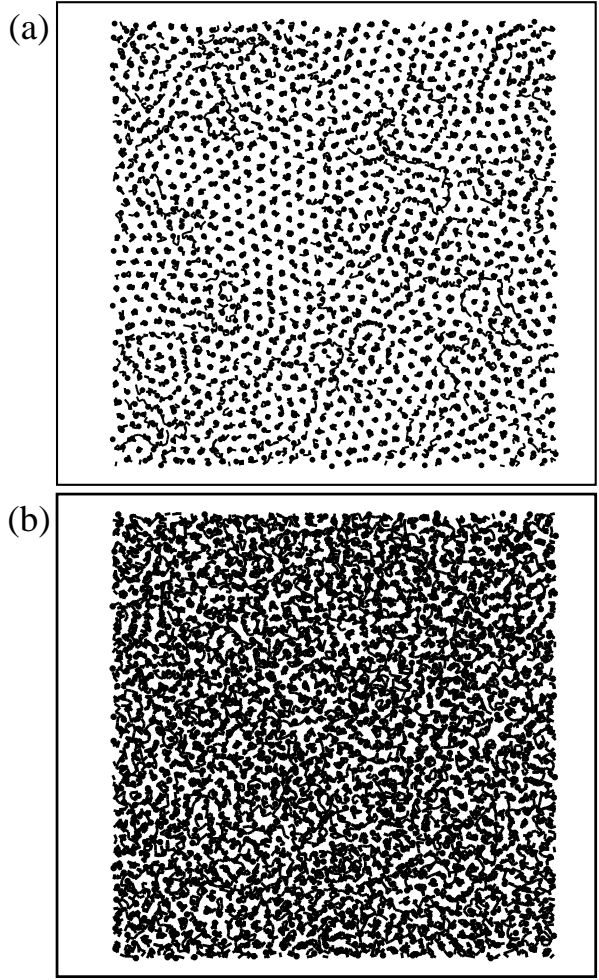}}
\caption{
Particle positions (black dots) and trajectories (black lines) for 
a fixed period of time for (a) $T/T_{d} = 1.04$ and (b) $T/T_{d} = 7.0$.
}
\end{figure}

\hspace{-13pt}
there are no correlations in the
motion. 

We next consider a way to characterize the behaviors of the system 
at the different temperatures by measuring the fluctuations of the
defect density as a function of time. We compute the defect
configuration and density
every 20 time steps for 20000 frames
and obtain a time series of the defect density vs time
for several temperatures from $T/T_{d} = 1.04$ to $10$. 
We have
considered a variety of sampling rates and find consistent results.
In Fig.~3(a) we show a portion of the
time series of the fraction of $n_c=6$
particles, $P_6(t)$, 
for $T/T_{d} = 1.04$  (upper curve) and $T/T_{d} = 7.0$ (lower
curve). For $T/T_{d} = 1.04$ the fluctuations show long time variations,
while for the higher temperature the fluctuations are very rapid. 
In Fig.~3(b) we plot the power spectrum $S(f)$ 
of $P_6(t)$ for $T/T_{d} = 1.04$,
which fits well to a $1/f^{\alpha}$ scaling over more than three decades
with the best fit $\alpha = 1.04$, close to $1/f$ noise. 
As the temperature increases, the spectrum changes from $1/f$ to 
white noise ($\alpha = 0$) for low
frequencies. The small frequencies correspond to the long 

\begin{figure}
\center{
\epsfxsize=3.5in
\epsfbox{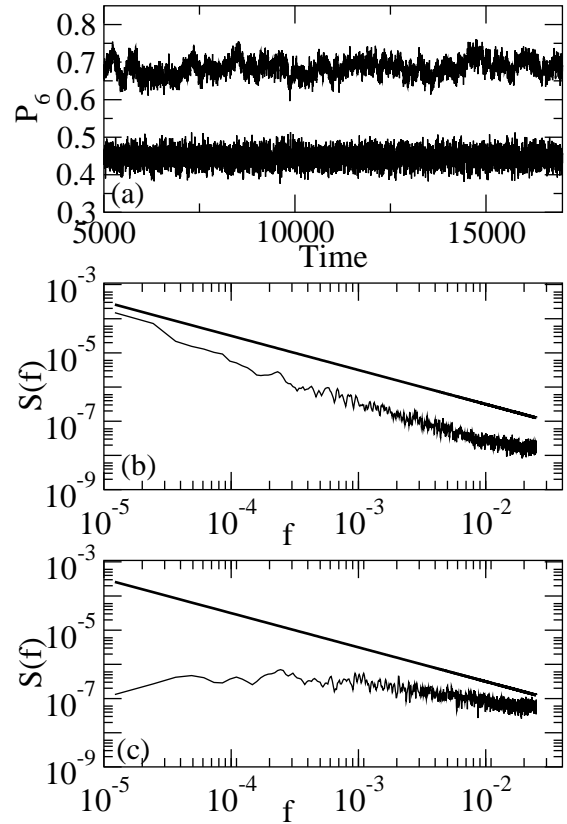}}
\caption{
(a) A portion of the time series of the 
fraction of $n_c=6$ particles, $P_{6}(t)$, for
(upper curve) $T/T_{d} = 1.04$ and (lower curve) $T/T_{d} = 7.0$.
(b) The power spectrum for the time series at $T/T_{d} = 1.04$.  The
solid line indicates a slope of $1/f$.
(c) The power spectrum for the time series at $T/T_{d} = 7.0$ along with
a $1/f$ line.
}
\end{figure}

\hspace{-13pt}
time 
correlations in the system indicating that correlation times of the
defect creation or annihilation decrease at higher temperatures. 
In Fig.~3(c), the power spectrum for $T/T_{d} = 7.0$ is
white with $\alpha = 0$ for a large portion of the curve.
We have also considered $P_6(t)$ for
$T/T_{d} < 1.0$ where there are a small number of bound
dislocation pairs present. Here we find white noise with
small amplitude fluctuations.

The magnitude of the noise power, 
obtained by integrating $S(f)$ over the first octave of frequencies,
is related to the defect density and
the disordering transition. 
In Fig.~4(a) we show $P_6$ as a 
function of $T$ and in Fig.~4(b) we show the corresponding noise power 
$S_{0}$. The peak in $S_{0}$ coincides with the onset of the
defect proliferation. For increasing $T$, the noise power falls 
and saturates when the spectrum becomes
white. There is a finite $S_{0}$ for 
$T/T_{d} < 1.0$ since, as noted, pairs of bound dislocations
can still be thermally created. 
We have repeated the same set of simulations for particles interacting
with a long range logarithmic potential and again find a
noise power peak coinciding with 
the defect proliferation 

\begin{figure}
\center{
\epsfxsize=3.5in
\epsfbox{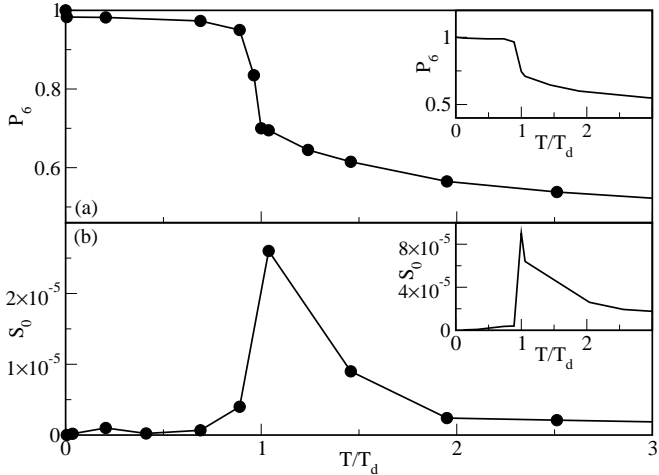}}
\caption{
(a) The fraction of six-fold particles $P_{6}$ vs $T$ for a system 
of particles with a screened Coulomb interaction. Inset: 
$P_{6}$ vs $T$ for a system with logarithmic interactions.
(b) The integrated noise power $S_{0}$, obtained from $S(f)$,
vs $T$ for the same system in (a). 
Inset: $S_{0}$ vs $T$ for a system with logarithmic interactions. 
}
\end{figure}

\hspace{-13pt}
(insets of Fig.~4) and a $1/f$ power spectrum
just above $T_{d}$. 

We have also considered a sudden quench from $T/T_{d} > 1.0$
to $T/T_{d} < 1.0$.  Here the system is out of equilibrium, 
and approaches equilibrium by means of the annihilation of
the defects.  In Fig. 5 we show the particle motions after a quench
from $T/T_{d} = 1.04$ to $T/T_{d} = 0.2$. 
We find that the number of defects decreases
rapidly to a saturation point where a few defects or grain 
boundaries remain. The motion of the particles corresponds to the 
annihilation of the defects. As seen in Fig.~5,
the motion during the defect annihilation process 
has the same string like nature
as the equilibrium $T>T_d$ 
motion near the defect proliferation transition. 
This result suggests that the heterogeneous particle motion 
is produced by the motion of defects,
particularly the creation or annihilation of these defects in a correlated
manner.    

In summary, we have proposed a new way to examine  dynamical heterogeneities  
in a liquid 
by measuring the fluctuations in the topological defect
density. We find that near the onset of defect proliferation, the
particle motion is heterogeneous and the defects cluster together into
grain boundaries or strings. The defect fluctuations in this phase have
a $1/f$ character and a large noise power. 
For higher temperatures the heterogeneities are lost and
the fluctuation spectrum is white. We have also examined the defect
annihilation after a quench from the disordered phase to the ordered phase
and find that the particle motion also shows heterogeneities as
the defects are annihilated.
We have considered systems with screened Coulomb interactions and
logarithmic interactions and find similar behavior.
 Our predictions can be easily tested in 
2D dense colloidal liquids.
It would also 

\begin{figure}
\center{
\epsfxsize=3.5in
\epsfbox{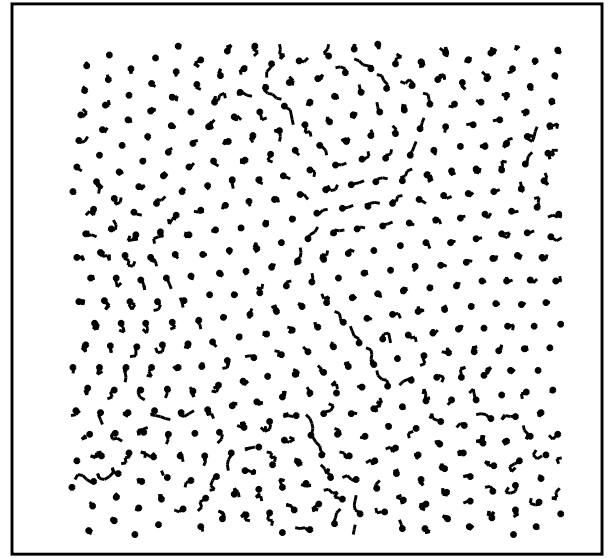}}
\caption{
The particle positions (black dots) and trajectories (black lines) for
a system 
quenched from $T/T_{d} = 1.04$ to $T/T_{d} = 0.3$. 
The
trajectories are analyzed once the system is at the lower temperature. 
During this time a portion of the defects are annihilated. 
 }
\end{figure}

\hspace{-13pt}
be interesting to study the
fluctuations of defects in 3D. 
This would be experimentally
possible in 3D colloidal assemblies using confocal microscopy.   

We acknowledge helpful discussions with S. Bhattacharya, D. Grier, 
V. Vinokur, E. Weeks, and M. Weissman.
This work was supported by the US Department of Energy under Contract No.
W-7405-ENG-36.


\begin{references}

\bibitem{Liquieds}
For reviews, see: 
S.C.~Glotzer, J. Non-Crystalline Solids {\bf 274}, 342 (2000);
R.~Richert, J. Phys. Condens. Mat. {\bf 14}, R703 (2002). 

\bibitem{Kob}
W.~Kob {\it et al.}, Phys.~Rev.~Lett.~{\bf 79}, 2827 (1997);
C.~Donati {\it et al.}, Phys.~Rev.~Lett.~{\bf 80}, 2338 (1998);
M.M.~Hurley and P.~Harrowell, Phys.~Rev.~E {\bf 52}, 1694 (1995).

\bibitem{Kegel}
W.K.~Kegel and A.~van Blaarderen, Science {\bf 287}, 290 (2000);
E.R.~Weeks {\it et al.}, Science {\bf 287}, 627 (2000). 

\bibitem{Cui}
B.~Cui, B.~Lin, and S.A.~Rice, J.~Chem.~Phys.~{\bf 114}, 9142 (2001).

\bibitem{Tang}
Y.~Tang {\it et al.}, Phys.~Rev.~Lett.~{\bf 62}, 2401 (1988).

\bibitem{Marcus}
A.H.~Marcus, J.~Schofield, and S.A.~Rice, Phys.~Rev.~E {\bf 60}, 5725 (1999).

\bibitem{Goree}
R.A.~Quinn and J.~Goree, Phys.~Rev.~E {\bf 64}, 051404 (2001).

\bibitem{Chiang}
C.-H. Chiang and Lin I, Phys.~Rev.~Lett.~{\bf 77}, 647 (1996).

\bibitem{Tau}
W.T. Juan and Lin I, Phys.~Rev.~Lett.~{\bf 80}, 3073 (1998).

\bibitem{Stranburg}
K.J.~Stranburg, Rev.~Mod.~Phys.~{\bf 60}, 161 (1988).

\bibitem{Jensen}
N.~Gr{\o}nbech-Jensen {\it et al.}, Mol.~Phys.~{\bf 92}, 941 (1997).

\bibitem{Canright}
F.L. Somer {\it et al.}, Phys.~Rev.~Lett.~{\bf 79}, 3431 (1997).

\end{references}
\end{document}